\documentclass[reqno]{amsart}
\usepackage{verbatim}
\newcommand{\heute}{23. May 2003}
\newtheorem{thm}{Theorem}

\newtheorem{lemma}{Lemma}
\theoremstyle{definition}

\newtheorem{hyp}{Hypothesis}
\newtheorem{dfn}{Definition}
\newcommand{\addsupport}[1]{} 
\newcommand{\extra}[1]{} 
\newcommand{\bb}[1]{{\mathbb{#1}}}
\newcommand{\e}{\hbox{\rm e}}
\newcommand{\res}{\operatorname{res}}

\renewcommand{\Im}{\operatorname{Im}}
\renewcommand{\Re}{\operatorname{Re}}
\renewcommand{\P}{\mathcal P}
\begin{document}
\title[The inverse resonance problem]{The inverse resonance problem for
perturbations of algebro-geometric potentials}

\author[Brown, Weikard]{B. M. Brown and R. Weikard}

\address{Department of Computer Science, University of Wales, Cardiff,
PO Box 916, Cardiff CF2 3XF, U.K.}
\email{Malcolm.Brown@cs.cf.ac.uk}

\address{Department of Mathematics, University of Alabama at
Birmingham, Birmingham, Alabama 35294-1170, USA}
\email{rudi@math.uab.edu}

\thanks{}

\date{\heute}
\maketitle

\bibliographystyle{plain}

\section{Introduction}
Algebro-geometric potentials $q$ of the one-dimensional Schr\"odinger
operator $L=-d^2/dx^2+q$ are a small but important class of potentials
with a number of rather nice properties. Their name is due to the fact
that there exists an a differential operator $P$ of odd order such that
the pair $(P,L)$ satisfies the algebraic relation $P^2=R(L)$ for some
polynomial $R$ of odd degree. Moreover, the commutator $[P,L]$ equals
zero, so that, according to Lax's celebrated paper \cite{MR38:3620},
$q$ is a stationary solution of some equation in the KdV hierarchy.
However, the property which is most important for this paper is that
the the solutions of the differential equation $-y''+qy=\lambda y$ are
meromorphic functions of the independent variable (see Its and Matveev
\cite{MR57:18570} and Segal and Wilson
\cite{MR87b:58039}).\footnote{That meromorphic solutions of the
differential equation are also a sufficient condition for $q$ to be
algebro-geometric in at least the elliptic, simply periodic, and
rational realm was shown in \cite{MR97f:14046} and
\cite{MR2000d:35208}. See \cite{MR99i:58075} for an overview of the
subject.} At least in some subclasses of algebro-geometric potentials
this property allows to show that solutions have a certain algebraic
structure (see Theorem \ref{strct}). Asymptotic features of this
structure survive when the potential is subjected to compactly
supported perturbations. The paradigm of an algebro-geometric potential
is the $g$-soliton $-g(g+1)/\cosh(x)^2$ (and the functions obtained
from it under time evolution according to the KdV equation) but the set
includes also rational, periodic, elliptic and other functions.

V. A. Marchenko \cite{Marchenko1955} showed in 1955 that a real-valued
potential $q$ on $[0,\infty)$ for which $(1+x)q(x)$ is integrable is
uniquely determined from the scattering phase, the eigenvalues and
their norming constants.

Neither the scattering phase (as a function on $\bb R$) nor the norming
constants can be obtained directly from laboratory measurements and one
might therefore ask what would constitute equivalent information. In
the case of compactly supported potentials at least the answer is also
(implicitly) given by Marchenko: Each piece of the required information
can be obtained from the Jost function\footnote{The solution
$\psi(z,\cdot)$ of $-y''+qy=z^2y$ which asymptotically equals $\exp(i z
x)$ is called the Jost solution of the problem; the function
$\psi(\cdot,0)$ is then called the Jost function.} which, in this
particular case, is an entire function. Thus it can be recovered from
the location of its zeros (and its known asymptotic behavior along the
positive imaginary line) with the aid of Hadamard's factorization
theorem. From a physical point of view the squares of these zeros are
(Dirichlet) eigenvalues or resonances (depending on whether the zero in
question is in the upper or lower half plane). In other words, the
location of eigenvalues and resonances determines a compactly supported
real-valued potential.

It is now important to realize that both eigenvalues and resonances
appear as poles of the scattering amplitude. In contrast to eigenvalues
the resonance poles are complex but they still produce a large bump in
the scattering cross section for real energies if they are close enough
to the real axis. Since the scattering cross section is measured in the
laboratory the observation that eigenvalues and resonances determine
the potential is very important from a practical point of view. To our
knowledge this observation was first publicly made by Korotyaev
\cite{korotyaev} but Zworski \cite{MR1856251} had realized (but not
published) it earlier in the context of compactly supported even
potentials on $\bb R$. Christiansen \cite{TC} is another recent
contribution to this field.

In joint work with I. Knowles we have developed an approach to this
kind of problems which is independent from Marchenko's (see
\cite{BKW}). Our approach makes no distinction whether $q$ is real or
complex by reconstructing the Weyl-Titchmarsh $m$-function instead the
scattering phase. The $m$-function may be defined for complex
potentials (see Brown et. al. \cite{MR2000c:34063}) and determines the
potential uniquely, just as in the case of real potentials (see
\cite{BPW}). In this paper we will apply this new method to compactly
supported perturbations of algebro-geometric potentials. Note that
these potentials generally do not have an integrable first moment so
that Marchenko's approach does not apply.

In Section \ref{prelim} we present some basic facts about the
Tichmarsh-Weyl $m$-function for complex-valued potentials. Section
\ref{mainproof} states and proves a general theorem (Theorem
\ref{main}) which identifies sufficient conditions under which the
eigenvalues and resonances determine a potential. Algebro-geometric
potentials are defined and discussed in Section \ref{ag}. In
particular, we show there that they provide examples for Theorem
\ref{main}. Finally, in Section \ref{csp} we discuss compactly
supported perturbations of algebro-geometric potentials. Theorem
\ref{example} is the main result of that section.

\section{Preliminaries} \label{prelim}
Let $\Sigma$ be a fixed open sector of the complex plane whose vertex
is at the origin. Then define $\mathcal Q_\Sigma$ to be the set of
those complex-valued, locally integrable functions on $[0,\infty)$ for
which there is an open half plane $\Lambda$ satisfying the following
two requirements:\footnote{If $S$ is a subset of the complex plane we
denote its complement by $S^c$ and its closed convex hull by
$\overline{\rm co}(S)$.}
\begin{enumerate}
\item $\Lambda^c\cap\Sigma$ is bounded.
\item The set $Q=\overline{\rm co}(\left\{q(x)+r:
x,r\in[0,\infty)\right\})$ does not intersect $\Lambda$.
\end{enumerate}

Conditions of this type have first been introduced by Brown et al.
\cite{MR2000c:34063}. Given a function $q\in\mathcal Q_\Sigma$ we
consider the differential expression $L=-d^2/dx^2+q$ on $[0,\infty)$.
We will say that $q$ is of Class I, if at most one (up to constant
multiples) solution of $L y=\lambda y$ is square integrable on
$[0,\infty)$. Otherwise, if all solutions of $L y=\lambda y$ are square
integrable on $[0,\infty)$, we will say that $q$ is of Class II. This
classification is independent of the choice of $\lambda$. For
real-valued potentials it coincides with the classical limit-point and
limit-circle distinction. However, for complex-valued potentials it
does not coincide with Sims's distinction (cf. \cite{MR19:144g})
between the limit-point and limit-circle cases. See \cite{BPW} for a
discussion of this issue.

Now let $\theta(\lambda,\cdot)$ and $\phi(\lambda,\cdot)$ be linearly
independent solutions of $Ly=\lambda y$ satisfying the initial
conditions\footnote{Throughout the paper we will use the following
notation for derivatives: If $f$ is a function of several variables we
will use $\dot f$ and $f'$ to denote the derivative of $f$ with respect
to the first and last variable, respectively. If $f$ is a function of
two variables $f^{(j,k)}$ denotes the function obtained by
differentiating $j$ times with respect to the first variable and $k$
times with respect to the second.}
\begin{eqnarray*}
 \theta(\lambda,0) = 1 & & \phi(\lambda,0) = 0\\
 \theta'(\lambda,0) = 0 & & \phi'(\lambda,0)= 1.
\end{eqnarray*}
It is shown in Brown et al. \cite{MR2000c:34063} (see also \cite{BPW})
that for every $\lambda\in\Lambda$ there is at least one square
integrable solution of $Ly=\lambda y$ which is not a multiple of
$\phi(\lambda,\cdot)$. Hence, if $q$ is of Class I and
$\lambda\in\Lambda$, then there is precisely one square integrable
solution $\chi(\lambda,\cdot)$ (up to constant multiples) and there is
a unique number $m(\lambda)$ such that $\chi(\lambda,\cdot)
=\theta(\lambda,\cdot)+m(\lambda)\phi(\lambda,\cdot)$ is square
integrable. This function $m:\Lambda\to \bb C:\lambda\mapsto
m(\lambda)$ is the generalization of the Titchmarsh-Weyl $m$-function
for a Dirichlet boundary condition at zero to the case of
complex-valued potentials. Note that
$$m(\lambda)=\frac{\chi'(\lambda,0)}{\chi(\lambda,0)}.$$

Just as in the selfadjoint case $m$ is an analytic function (see
\cite{MR2000c:34063}). It may well be possible to extend it
analytically to a larger domain than $\Lambda$. Sometimes $m$ may even
be extended to the Riemann surface of $\lambda\mapsto\sqrt{\lambda}$.
This is the case we are interested in and therefore we introduce the
function
$$M(z)=m(z^2)$$
putting the branch cut on the positive real axis (so that $\Im(z)>0$
represents the so called physical $\lambda$-sheet).

\section{The main theorem} \label{mainproof}
\begin{dfn} \label{dcw}
Given an odd polynomial $W\in\bb C[z]$ of degree $2g+1$ define
$\mathcal C_{W}$ to be the family of potentials $q\in\mathcal Q_\Sigma$
which are of Class I and for which there exist functions $\psi:\bb
C\times[0,\infty)\to\bb C$ satisfying the following conditions:
\begin{enumerate}
\item For every complex number $z$ the functions $\psi(z,\cdot)$ and
$\psi(-z,\cdot)$ are nontrivial solutions of the differential equation
$-y''+qy=z^2 y$.

\item The Wronskian of $\psi(z,\cdot)$ and $\psi(-z,\cdot)$ satisfies
$$\psi(z,\cdot)\psi'(-z,\cdot)-\psi(-z,\cdot)\psi'(z,\cdot)=W(z).$$

\item $\psi(z,\cdot)$ is square integrable for all $z$ in some
nonempty open subset of the upper halfplane $\bb C_+$.

\item $\psi(\cdot,0)$ and $\psi'(\cdot,0)$ are entire functions of
finite growth order.

\item There exists a ray such that $\psi(z,0)/z^g$ tends to one as $z$
tends to infinity along the ray.

\item There is an integer $p$ and a sequence of circles $t\mapsto
r_n\exp(it)$ such that $r_n$ tends to infinity and
$|(\psi'/\psi)(r_n\exp(it),0)| r_n^{-p-1}$ tends to zero uniformly for
$t\in[0,2\pi]$.
\end{enumerate}
\end{dfn}

\begin{thm} \label{main}
Let $W\in\bb C[z]$ be an odd polynomial of degree $2g+1$ and assume
that $q$ is a potential in $\mathcal C_W$ with $\psi$ being the
function from Definition \ref{dcw} establishing that fact. If $W(z)=0$
implies $\psi(z,0)\neq0$ then the zeros of $\psi(\cdot,0)$ and their
multiplicities determine $q$ uniquely.
\end{thm}

\begin{proof}
It is well-known that, in the self-adjoint case, the Titchmarsh-Weyl
$m$-function determines the potential $q$ (see Bennewitz
\cite{MR2001m:34035} for a rather concise proof). It was shown in
\cite{BPW} that this remains true even if $q$ is complex-valued at
least as long as it is of Class I. Since, of course, $M$ determines
$m$, we only have to show that the given information suffices to
determine $M$.

It follows from condition (3) that
$$M(z)=\frac{\psi'(z,0)}{\psi(z,0)}.$$
Condition (4) implies that $M$ is meromorphic and that its poles are
the zeros of $\psi(\cdot,0)$. We denote the poles of $M$ by the
pairwise distinct numbers $z_1, z_2, ...$ and we use $n_1, n_2, ...$
for their respective multiplicities. The zeros are labelled such that
$|z_1|\leq |z_2| \leq ...$.

Let $h_z(\mu)=(z/\mu)^{p+1}/(z-\mu)$. Also define $\gamma_n(t)
=r_n\exp(it)$ for $t\in[0,2\pi]$ and $B_n=\{z:|z|<r_n\}$. Then, by the
residue theorem,
$$\frac1{2\pi i} \int_{\gamma_n} h_z(\mu)M(\mu)d\mu
 = -M(z)+\sum_{k=0}^p \frac{M^{(k)}(0)}{k!}z^k +
 \sum_{z_j \in B_n} \res_{z_j}(h_zM)$$
if $0\neq|z|<r_n$ and if $z$ is none of the poles of $M$. According to
condition (6) the integral on the left tends to zero as $n$ tends to
infinity proving firstly the convergence of the series and secondly
that
\begin{equation} \label{0324.1}
M(z)=\sum_{k=0}^p \frac{M^{(k)}(0)}{k!}z^k +
 \sum_{j=1}^\infty \res_{z_j}(h_zM).
\end{equation}

Suppose we had already determined the infinite series on the right
hand side of equation \eqref{0324.1}. We can then find the
polynomial $\sum_{k=0}^p M^{(k)}(0)z^k/k!$ from the asymptotic
behavior of the $m$-function along some ray since $m(z^2)=iz+o(1)$
as $z$ tends to infinity along certain rays (see Theorem 6 of
\cite{BKW}).

Thus the theorem is proved once we determine the residues of $h_z M$ at
the poles of $M$. To do this let
$$f_j(\mu)=\frac{(\mu-z_j)^{n_j}}{\psi(\mu,0)}.$$
Then
\begin{align*}
\res_{z_j}(h_zM)&=
 \frac1{(n_j-1)!} (\psi'(\cdot,0)h_zf_j)^{(n_j-1)}(z_j)\\
 &=\frac1{(n_j-1)!} \sum_{r=0}^{n_j-1}\binom{n_j-1}{r}
  \psi^{(r,1)}(z_j,0) (h_zf_j)^{(n_j-1-r)}(z_j)
\end{align*}
and this quantity may be computed once we know the function
$\psi(\cdot,0)$ (and hence the functions $f_j$) and the numbers
$\psi^{(r,1)}(z_j,0)$ for $r=0, ..., n_j-1$. We will now show that this
information can be obtained from the given data.

Firstly, $\psi(\cdot,0)$ is given through Hadamard's factorization
theorem as
$$\psi(z,0)=z^k\exp(g(z)) \prod_{n=1}^\infty E_\rho(z/z_n)$$
where $k$ and $\rho$ are integers, $g$ is a polynomial, and
$$E_\rho(z)=(1-z)\exp(z+\frac{z^2}2+...+\frac{z^\rho}\rho).$$
The number $\rho$ is to be chosen such that $\sum_{j=1}^\infty n_j
|z_j|^{-\rho+1}$ is finite. This is always possible since otherwise
$\psi(\cdot,0)$ would not have finite growth order. The polynomial $g$
may be determined from the asymptotic behavior of $\psi(\cdot,0)$,
given in condition (5), and we have $k=0$ since $\psi(0,0)\neq0$ (zero
is a root of $W$).

Secondly, taking $r$ derivatives of the equation
$$\psi(z,\cdot)\psi'(-z,\cdot)-\psi(-z,\cdot)\psi'(z,\cdot)=W(z)$$
with respect to $z$ and evaluating them at $z_j$ gives that
$$\psi^{(r,1)}(z_j,0) \psi(-z_j,0)=-W^{(r)}(z_j)
 -\sum_{s=0}^{r-1} \frac{(-1)^{r-n}r!}{(r-n)!n!}
  \psi^{(s,1)}(z_j,0) \psi^{(r-s,0)}(-z_j,0)$$
as long as $r\leq n_j-1$ since $z_j$ is a zero of
$\psi(\cdot,0)=0$ of order $n_j$. We know that $\psi(-z_j,0)\neq0$
since $\psi(z_j,\cdot)$ and $\psi(-z_j,\cdot)$ are linearly
independent. Hence the numbers $\psi^{(0,1)}(z_j,0)$, ...,
$\psi^{(n_j-1,1)}(z_j,0)$ may be computed recursively.
\end{proof}

\section{Algebro-geometric potentials} \label{ag}
Let $L$ be the differential expression $L=-d^2/dx^2+q$. A meromorphic
function $q$ is called algebro-geometric (or an algebro-geometric
potential) if there exists an ordinary differential expression $P$ of
odd order which commutes with $L$. The reason behind this choice of
words is that, according to results of Burchnall and Chaundy
\cite{BC1}, \cite{BC2}, the differential expressions $P$ and $L$
commute if and only if there exists a polynomial $\mathcal Q$ in two
variables such that $\mathcal Q(P,L)=0$.

In this section we consider potentials which are either rational
functions or else simply periodic meromorphic functions of period $p$
bounded at the ends of the period strip. We define
$$\xi(x)=\begin{cases}
 x&\text{in the rational case,}\\
 \frac{p}{2\pi i}(\e^{2\pi ix/p}-1)&\text{in the periodic case.}
\end{cases}$$
and
\begin{equation} \label{240303.1}
\P(x)=\frac{\xi'(x)}{\xi(x)^2}
\end{equation}
or explicitly
$$\P(x)=\begin{cases}
 1/x^2&\text{in the rational case,}\\
 \left(\frac{2\pi i}{p}\right)^2\e^{2\pi ix/p}/(\e^{2\pi ix/p}-1)^2
 &\text{in the periodic case.}
\end{cases}$$
Note that, as $p$ tends to infinity, $\xi(x)$ tends to $x$. It will be
convenient to refer to the rational case as the case where $p$ is
infinite.

Remarks:
\begin{enumerate}
\item $\P(x)$ tends to zero if $x$ tends to the ends of the period
strip (defining the whole complex plane as the period strip if $p$ is
infinite). Moreover, the principal part of $\P$ at zero equals $1/x^2$.

\item It was shown in \cite{MR2000d:35208} that the potential $q_0$ is
algebro-geometric if and only if all solutions of the equation
$-y''+q_0y=\lambda y$ are meromorphic for all complex numbers $\lambda$
provided that $q_0$ is either rational or else simply periodic,
meromorphic, and bounded at the ends of the period strip. This provides
a simply criterion by which one may determine whether a potential is
algebro-geometric.
\end{enumerate}

\begin{thm} \label{strct}
Suppose that $q_0$ is a rational function bounded at infinity or a
simply periodic meromorphic function bounded at the ends of the period
strip and that $q_0$ is algebro-geometric. Then the following
statements hold:
\begin{enumerate}
\item $q_0(x)=\lambda_0 + \sum_{j=1}^m s_j(s_j+1)\P(x-x_j)$ for a
suitable choice of the parameters $\lambda_0$, $m$, $s_1$, ..., $s_m$
(we let $m=0$ if $q_0$ is constant) and suitable pairwise distinct
(modulo the period in the periodic case) points $x_1, ..., x_m$.

\item There is a nonnegative integer $g$ and there are rational
functions $r_0,...,r_{g-1}$ such that $\psi_0(z,x)
=(z^g+r_{g-1}(\xi(x))z^{g-1}+...+r_0(\xi(x))) \exp(i z x)$ is a
solution of the equation $-y''+(q_0-\lambda_0)y=z^2 y$.

\item The Wronskian $W$ of $\psi_0(z,\cdot)$ and $\psi_0(-z,\cdot)$ is an
odd polynomial in $\bb C[z]$ of degree $2g+1$. In the rational case its
only zero is $z=0$, i.e., $W(z)=-2(iz)^{2g+1}$. In the periodic case
all zeros of $W$ are simple.
\end{enumerate}
\end{thm}

\begin{proof}
Statement (1) was proved in \cite{MR2000d:35208}. It was proved in
\cite{MR2001a:34146} that $-y''+q_0y=\lambda y$ has at least one
solution $\psi_0(z,\cdot)$ of the form given. It is then a
straightforward calculation to show that $\psi_0(-z,\cdot)$ also yields
a solution. This gives statement (2). Statement (3) was proved in
\cite{GUW}.
\end{proof}

These results enable us to show that $q_0$ is in $\mathcal C_W$ where
$W$ is the Wronskian just introduced: Theorem \ref{strct} shows the
validity of condition (1). Condition (2) is satisfied by definition. If
$q_0$ is rational then $\psi_0(z,\cdot)$ is square integrable for any
$z$ in the upper half plane. In the periodic case $\e^{izx}$ decays
faster than any power of $\xi(x)$ provided that $\Im(z)$ is
sufficiently large. Hence $\psi_0(z,\cdot)$ is square integrable for
any $z$ with sufficiently large imaginary part. This proves that
condition (3) holds. Obviously $\psi_0(\cdot,0)$ as well as
$\psi_0'(\cdot,0)$ are polynomials and hence entire functions of growth
order zero. Condition (5) holds for any ray in the complex plane. Since
$$M(z)=\frac{\psi'(z,0)}{\psi(z,0)}=iz+O(z^{-1})$$
condition (6) is satisfied with $p=1$. The circles may eventually be
chosen arbitrarily as $M$ has only finitely many poles.

\section{Compactly supported perturbations of base potentials} \label{csp}
\subsection{Transformation operators}
Throughout this section we require the following hypothesis to be
satisfied.
\begin{hyp} \label{H1}
$q$ and $q_0$ are locally integrable functions on $[0,\infty)$ and
there exists a positive number $R$ such that the support of $q-q_0$ is
contained in $[0,R]$. Moreover, the following estimate holds:
$$M=\sup\left\{\int_0^{(t-x)/2}|q(\alpha-\beta)-q_0(\alpha+\beta)|d\beta:
 \frac{t-x}{2}\leq \alpha \leq R, 0\leq x\leq t\right\}<\infty.$$
\end{hyp}

Let $\Omega=\{(t,x)\in\bb R^2: 0\leq x\leq t\}$ and
$\Omega_0=\{(t,x)\in\bb R^2: 0\leq x\leq t\leq 2R-x\}$. For
$(t,x)\in\Omega$ define
$$K_0(t,x)=\frac12 \int_{(t+x)/2}^\infty (q(s)-q_0(s))ds$$
and, for $n\in\bb N$,
$$K_n(t,x)=\int_{(t+x)/2}^\infty
 \int_0^{(t-x)/2} (q(\alpha-\beta)-q_0(\alpha+\beta))
 K_{n-1}(\alpha+\beta,\alpha-\beta) d\beta d\alpha.$$

\begin{lemma} \label{L5.1}
If $(q,q_0)$ satisfies Hypothesis \ref{H1} then, for every nonnegative
integer $n$,
$$|K_n(t,x)| \leq \frac12 \frac{M^n}{n!} \left(R-\frac{t+x}2\right)_+^n
 \int_{(t+x)/2}^R |q(s)-q_0(s)|ds.$$
In particular, $K_n(t,x)=0$ if $t+x\geq 2R$.
\end{lemma}

\begin{proof}
The lemma will proved by induction on $n$. The statement is true for
$n=0$. Assume now that it holds for $n-1$.

If $t+x\geq 2R$ then $K_{n-1}(\alpha-\beta,\alpha+\beta)=0$ for all
$(\alpha,\beta)$ in the domain of integration and hence $K_n(t,x)=0$.

If $t+x\leq 2R$ we obtain
\begin{align*}
|K_n(t,x)| &\leq \int_{(t+x)/2}^R \int_0^{(t-x)/2}
   |q(\alpha-\beta)-q_0(\alpha+\beta)||K_{n-1}(\alpha+\beta,\alpha-\beta)|
   d\beta d\alpha \\
&\leq \frac12 \frac{M^{n-1}}{(n-1)!} \int_{(t+x)/2}^R (R-\alpha)^{n-1}
   \int_{\alpha}^R |q(s)-q_0(s)|ds  M d\alpha \\
&\leq \frac12 \frac{M^n}{(n-1)!} \int_{(t+x)/2}^R |q(s)-q_0(s)|ds
\int_{(t+x)/2}^R (R-\alpha)^{n-1} d\alpha.
\end{align*}
\end{proof}

Because of this lemma the function
$$K(t,x)=\sum_{n=0}^\infty K_n(t,x)$$
is well defined for $0\leq x\leq t$ and satisfies the integral equation
\begin{align*}
K(t,x)=&\frac12 \int_{(t+x)/2}^R (q(s)-q_0(s))ds\\
&+\int_{(t+x)/2}^R
 \int_0^{t-x} (q(\alpha-\beta)-q_0(\alpha+\beta))
 K(\alpha+\beta,\alpha-\beta) d\beta d\alpha.
\end{align*}

We also define
$$H(t,x)=\int_{(t+x)/2}^R\int_0^{t-x}h(\alpha,\beta)d\beta d\alpha
 =\sum_{n=1}^\infty K_n(t,x)$$
where
$$h(\alpha,\beta)=(q(\alpha-\beta)-q_0(\alpha+\beta))
 K(\alpha+\beta,\alpha-\beta)$$
so that
$$K(t,x)=K_0(t,x) + H(t,x).$$

\begin{lemma} \label{L5.2}
If $(q,q_0)$ satisfies Hypothesis \ref{H1} the following statements
hold:\\
 (i) $K$ is continuous on $\Omega$ and $H$ is continuously
differentiable on $\Omega$.\\
 (ii) For $x\in[0,R]$ the functions $K(\cdot,x)$, $H^{(1,0)}(\cdot,x)$,
and $H^{(0,1)}(\cdot,x)$ are uniformly absolutely continuous on
$[x,\infty)$. For $t\in(0,\infty)$ the functions $K(t,\cdot)$,
$H^{(1,0)}(t,\cdot)$, and $H^{(0,1)}(t,\cdot)$ are uniformly absolutely
continuous on $[0,t]$. Moreover,
\begin{equation} \label{240203.1}
H^{(2,0)}(t,x)-H^{(0,2)}(t,x)=-(q(x)-q_0(t))K(x,t).
\end{equation}
\end{lemma}

\begin{proof}
If $f$ is an absolutely continuous function both $x\mapsto f(x+t)$ and
$t\mapsto f(x+t)$ are uniformly absolutely continuous. Hence this is
the case for $K_0(\cdot,x)$ and $K_0(t,\cdot)$. Next one proves by
induction that $K_n(\cdot,x)$ and $K_n(t,\cdot)$ are continuously
differentiable and that their derivatives converge uniformly.

Next note that
$$H^{(1,0)}(t,x)=-\frac12 \int_0^{(t-x)/2} h((t+x)/2,\beta)d\beta
 +\frac12 \int_{(t+x)/2}^R h(\alpha,(t-x)/2)d\alpha$$
and
$$H^{(0,1)}(t,x)=-\frac12 \int_0^{(t-x)/2} h((t+x)/2,\beta)d\beta
 -\frac12 \int_{(t+x)/2}^R h(\alpha,(t-x)/2)d\alpha.$$
Uniform absolute continuity of these functions is then shown directly
using the uniform absolute continuity of $K$. The last claim follows by
direct computation.
\end{proof}

\begin{thm} \label{t3}
Assume that Hypothesis \ref{H1} is satisfied. If $y_0$ is in the
maximal domain of $-d^2/dx^2+q_0$ then $y$, given by
$$y(x)=y_0(x)+\int_x^{2R} K(t,x)y_0(t) dt,$$
is in the maximal domain of $-d^2/dx^2+q$. Moreover, if $-y_0''+q_0y_0
=z^2 y_0$, then $-y''+qy=z^2 y$.
\end{thm}

\begin{proof}
Since $y$ is eventually equal to $y_0$ and continuously differentiable
we only have to show that $y'$ is locally absolutely continuous and
that $-y''+qy$ is locally square integrable.

One computes
$$y'(x)=y_0'(x)-K(x,x)y_0(x)+\int_x^{2R} K_0^{(0,1)}(t,x)y_0(t)dt
 +\int_x^{2R} H^{(0,1)}(t,x)y_0(t)dt.$$
The preceding lemma shows that the last term on the right is locally
absolutely continuous. The substitution $t+x=2u$ shows that the third
term on the right is also locally absolutely continuous and that its
derivative is
$$\frac12 (q-q_0)(x) y_0(x)-\int_x^{2R} K_0^{(0,1)}(t,x)y_0'(t)dt.$$
Therefore
\begin{align*}
y''(x)=&\ y_0''(x)-K(x,x)y_0'(x)-(K^{(1,0)}+K^{(0,1)})(x,x)y_0(x)
 +\frac12 (q-q_0)(x) y_0(x)\\
 &\ -\int_x^{2R} K_0^{(0,1)}(t,x)y_0'(t)dt
 - H^{(0,1)}(x,x)y_0(x)+\int_x^{2R} H^{(0,2)}(t,x)y_0(t)dt.
\end{align*}
Two integrations by parts show that
\begin{align*}
\int_x^{2R} K_0^{(0,1)}(t,x)y_0''(t)dt =&\ -K(x,x)y_0'(x)
 +H^{(1,0)}(x,x)y_0(x)\\
 &\ -\int_x^{2R} K_0^{(1,0)}(t,x)y_0'(t)dt
 +\int_x^{2R} H^{(2,0)}(t,x)y_0(t)dt.
\end{align*}
Using that $K_0^{(1,0)}=K_0^{(0,1)}$ and equation \eqref{240203.1}
yields
$$-y''(x)+q(x) y(x)=-y_0''(x)+q_0(x)y_0(x)
 +\int_x^{2R} K(t,x)(-y_0''(t)+q_0(t)y(t))dt$$
which completes the proof of the theorem.
\end{proof}

\begin{hyp} \label{H2}
For some nonnegative integer $n$ the functions $q$ and $q_0$ have the
properties:\\
 (i) $q\in AC^n([0,R])$ and $q_0\in AC^n([0,2R])$.\\
 (ii) $q^{(j)}(R)=q_0^{(j)}(R)$ for $j=0,...,n-1$ but
$q^{(n)}(R)\neq q_0^{(n)}(R)$.
\end{hyp}

\begin{lemma} \label{L5.3}
Suppose that the hypotheses \ref{H1} and \ref{H2} are satisfied. Then
the following statements hold:\\
 (i) $K\in C^n(\Omega)$ and $K\in C^{n+1}(\Omega_0)$.\\
 (ii) $K^{(n+1,0)}(2R,0)=-(q-q_0)^{(n)}(R)/2^{n+2}\neq0$.\\
 (iii) If $(t,x)\in\Omega_0$ and $k+\ell=n+1$ then
$K^{(k,\ell)}(\cdot,x)\in AC([x,2R-x])$ and $K^{(k,\ell)}(t,\cdot)\in
AC([0,t])$.
\end{lemma}

\begin{proof}
Since $q$ and $q_0$ are $n$ times continuously differentiable every
derivative of order $r$ of $\int_{(t+x)/2}^R (q-q_0)(s)ds$ is given by
$$-\frac1{2^r} (q-q_0)^{(r-1)}((t+x)/2)$$
provided that $1\leq r\leq n+1$ and that $(t,x)\in\Omega_0$. We
therefore have to investigate only the function $H$. Let
$\Omega_0'=\{(\alpha,\beta)\in\bb R^2:0\leq\beta\leq\alpha\leq R\}$.
Induction shows that $h\in C^{r-1}(\Omega_0')$ implies that there are
numbers $a_{k,\ell,j}$, $b_{k,\ell}$, and $c_{k,\ell}$ such that
\begin{align}
 H^{(k,\ell)}(t,x)=&\sum_{j=0}^{k+\ell-2} a_{k,\ell,j}
 h^{(j,k+\ell-2-j)}(\frac{t+x}{2},\frac{t-x}{2})\nonumber \\
 &+b_{k,\ell}\int_0^{(t-x)/2}h^{(k+\ell-1,0)}(\frac{t+x}{2},\beta)
 d\beta\nonumber \\
 &+c_{k,\ell}\int_{(t+x)/2}^R h^{(0,k+\ell-1)}(\alpha,\frac{t-x}{2})
 d\alpha \label{260203.1}
\end{align}
provided that $1\leq k+\ell=r$. On the other hand
\begin{equation} \label{260203.2}
h^{(k,\ell)}(\alpha,\beta)
 =\sum_{\substack{\nu,\mu\geq 0\\ \nu+\mu\leq k+\ell}}
 g_{k,\ell,\nu,\mu} K^{(\nu,\mu)}(\alpha+\beta,\alpha-\beta)
\end{equation}
where
$$g_{k,\ell,\nu,\mu}=
 \tilde a_{k,\ell,\nu,\mu}q^{(k+\ell-\nu-\mu)}(\alpha-\beta)
 +\tilde b_{k,\ell,\nu,\mu}q_0^{(k+\ell-\nu-\mu)}(\alpha+\beta)$$
for certain numbers $\tilde a_{k,\ell,\nu,\mu}$ and $\tilde
b_{k,\ell,\nu,\mu}$. Hence $g\in C^r(\Omega_0')$ as long as $k+l=r\leq
n$. This shows that $K\in C^{n+1}(\Omega_0)$. We know already that $K$
is identically zero outside $\Omega_0$. To show that $K\in C^n(\Omega)$
we need $K^{(k,\ell)}(t,x)=0$ when $t+x=2R$ and $k+\ell\leq n$. But
\begin{equation} \label{260203.3}
K^{(k,\ell)}(t,x)=-\frac1{2^{k+\ell+1}}(q-q_0)^{(k+\ell-1)}((t+x)/2)
 +H^{(k,\ell)}(t,x).
\end{equation}
If $t+x=2R$ then the first term on the right is zero if $k+\ell\leq n$.
Using \eqref{260203.1} and \eqref{260203.2}, the second term is
expressed as a sum of terms involving expressions of the form
$K^{(r,s)}(R+\beta,R-\beta)$ where $r+s\leq k+\ell-1$ so that one can
show inductively that $H^{(k,\ell)}(t,x)=0$ when $t+x=2R$ even if
$k+\ell=n+1$. This completes the proof of (i) and (ii).

To prove (iii) we have to investigate \eqref{260203.3} again but for
variable $t$ (or $x$) and $k+\ell=n+1$. The first term is absolutely
continuous with respect to either variable in the stated intervals by
assumption. The second term is treated in the same way as $H^{(1,0)}$
or $H^{(0,1)}$ were treated in Lemma \ref{L5.2}.
\end{proof}

\subsection{The location of the resonances}
In this section we want to study the asymptotic location of resonances
for compactly supported perturbations of algebro-geometric potentials.
Therefore we assume henceforth the validity of the following
hypothesis.

\begin{hyp} \label{H3}
$q_0$ is an algebro-geometric potential of the form
$$q_0(x)=\lambda_0 + \sum_{j=1}^m s_j(s_j+1)\P(x-x_j)$$
where $\P$ is the function defined in \eqref{240303.1} and where
$\{x_j+np: n\in\bb Z, j=1,...,m\} \cap[0,\infty)=\emptyset$ (for
infinite $p$ this just means $\{x_j: j=1,...,m\}\cap[0,\infty)
=\emptyset$).

$q$ is a perturbation of $q_0$ such that $(q,q_0)$ satisfies hypotheses
\ref{H1} and \ref{H2}.
\end{hyp}

Under this hypothesis $q_0$ is real analytic on $[0,\infty)$. Moreover,
the functions $r_j$ introduced in Theorem \ref{strct} satisfy the
estimate
$$|r_j(\xi(x))| \leq \rho$$
for some number $\rho$ independent of $j\in\{0,..., g-1\}$ and
$x\in[0,2R]$.

Recall that
$$\psi_0(z,x)
 =(z^g+r_{g-1}(\xi(x))z^{g-1}+...+r_0(\xi(x)))\exp(i z x)$$
and define
$$\varphi(z,x)=\int_x^{2R} K(t,x)\psi_0(z,t) dt
 =\sum_{j=0}^g z^j \int_x^{2R} K(t,x)r_j(\xi(t))\exp(izt)dt$$
and
$$\psi(z,x)=\psi_0(z,x)+\varphi(z,x).$$

\begin{lemma} \label{L3a}
Suppose $(q,q_0)$ satisfies Hypothesis \ref{H3}. Then the following
statements hold:
\begin{enumerate}
\item \label{L3.5} Let $\Im(z)$ be fixed. Then $z^{-g}\varphi(z,x)$
tends to zero as $\Re(z)$ tends to $\pm\infty$.

\item \label{L3.6} $z^{-g}\varphi(z,x)$ tends to zero uniformly in
$\Re(z)$ as $\Im(z)\geq0$ tends to $\infty$.
\end{enumerate}
\end{lemma}

\begin{proof}
By the Riemann-Lebesgue lemma $\int_0^{2R} K(t,0)
r_j(\xi(t))\exp(izt)dt$ tends to zero as $\Re(z)$ tends to infinity.
This proves Statement \eqref{L3.5}. To prove \eqref{L3.6} note that
$$|z^{-g}\varphi(z,x)| \leq \sum_{j=0}^g |z|^{j-g}
 \int_0^{2R}|K(t,0)||r_j(\xi(t))|\e^{-\Im(z)t}dt
 \leq \frac{\rho\e^{MR}}{2\Im(z)} \|q-q_0\|_1$$
if $|z|\geq 1$.
\end{proof}

\begin{lemma} \label{L5}
Let $\nu$ be a positive number and $c_1$ a non-zero complex number.
Suppose that
$$\varphi(z,0)=\int_0^{2R} K(t,0)\psi_0(z,t) dt
 =z^g(c_1z^{-\nu} \e^{2izR}(1+f_1(z))+f_2(z))$$
where $|f_1(z)|\leq 1/12$ and $|f_2(z)|\leq1/6$ for all sufficiently
large $z$ in the closed lower half plane $\Im(z)\leq 0$. Then there is
a number $\tau$ such that $|\psi(z,0)|\geq|z|^g/3$ for all $z$ on the
semicircles given by $|z|=(2n\pi+\tau)/(2R)$ and $\Im(z)\leq 0$ and
sufficiently large integers $n$.
\end{lemma}

\begin{proof}
Recall that $\psi(z,0)=\psi_0(z,0)+\varphi(z,0)$ and that
$|z^{-g}\psi_0(z,0)-1|\leq 1/12$ when $|z|$ is sufficiently large.

We write $x=\Re(z)$, $y=\Im(z)$, and $c_1=\e^{\sigma+i\kappa}$ where
$\sigma,\kappa\in\bb R$. To prove the lemma we distinguish three cases.

First case: $|\Im(z)|\leq (\nu\log(n\pi/R)-\sigma-2)/(2R)$:\\ In this
case $\varphi(z,0)$ is negligible as compared to $z^g$ since
$$\left|c_1 z^{-\nu}\e^{2izR}\right|
 =\e^{\sigma-2Ry-2\log(n\pi/R)}(1+\frac{\tau}{2n\pi})^{-\nu}\leq1/6$$
for sufficiently large $n$. Hence $|z^{-g}\varphi(z,0)|\leq 25/72\leq
5/12$ but $|z^{-g}\psi_0(z,0)|\geq 11/12$ so that
$|z^{-g}\psi(z,0)|\geq 1/2$.

Second case: $|\Im(z)|\geq (\nu\log(n\pi/R)-\sigma+1)/(2R)$:\\ Here the
main contribution comes from the term $c_1 z^{-\nu}\e^{2izR}$. In fact,
when $n$ is sufficiently large,
$$\left|c_1z^{-\nu}\e^{2izR}(1+f_1(z))\right| \geq \frac{11}{6}
 \geq \frac74$$
while $\left|z^{-g}\psi_0(z,0)+f_2(z)\right| \leq 5/4$ so that
$|z^{-g}\psi(z,0)|\geq 1/2$.

Third case: $(\nu\log(n\pi/R)-\sigma-2)/(2R)\leq |\Im(z)|\leq
(\nu\log(n\pi/R)-\sigma+1)/(2R)$:\\ We obtain firstly that
$$|z^{-g}\psi(z,0)|
 \geq \left|1+c_1z^{-\nu}\e^{2izR}\right|-\frac1{12}-\frac3{12}-\frac16
 \geq \frac12+\Re\left(c_1z^{-\nu}\e^{2izR}\right)$$
since $|c_1z^{-\nu}\e^{2izR}|\leq 3$ when $n$ is sufficiently large.

Now let $\beta=\arg\left(c_1z^{-\nu}\e^{2izR}\right)
=2Rx+\kappa-\nu\arg(z)$ and note that $\arg(z)=3\pi/2 \pm \pi/2
+\arctan(y/x)$ where one has to choose the positive sign for positive
$x$ and the negative sign for negative $x$ (recall that $y$ is negative
in any case). After a small calculation one finds that $\pm
2Rx=2n\pi+\tau+r(n)$ where $r(n)=O(\log(n)^2/n)$ as $n$ tends to
infinity. This implies that $\arctan(y/x)=O(\log(n)/n)$ as $n$ tends to
infinity. Hence
\begin{align*}
\cos(\beta)&=\cos\left(\kappa+\frac{3\nu\pi}{2}
 \pm(\tau+\frac{\nu\pi}{2}+r(n))-\nu\arctan(y/x)\right) \\
 &\geq -|\sin(\pm r(n)-\nu\arctan(y/x))| \geq -\frac1{18}
\end{align*}
provided that $\tau$ is chosen in such a way that
$\cos(\kappa+3\nu\pi/2 \pm(\tau+\nu\pi/2))$ is nonnegative for either
choice of the sign. This can be achieved by choosing $\tau$ such that
$\tau+\nu\pi/2$ equals zero or $\pi$ depending on whether
$\cos(\kappa+3\nu\pi/2)$ is nonnegative or not. Therefore we arrive at
the following estimate
$$\Re\left(c_1z^{-\nu}\e^{2izR}\right)
 \geq -3|\sin(\pm r(n)-\nu\arctan(y/x))|\geq -\frac16$$
which holds for sufficiently large $n$.
\end{proof}

Note that the estimates in the above proof were made to show that the
circles can be chosen to avoid the resonances which are asymptotically
located near the points where $1+c_1z^{-\nu}\e^{2izR}=0$.

We will now show that the hypotheses of Lemma \ref{L5} can indeed be
satisfied under the hypotheses we made earlier.

\begin{lemma} \label{L7}
Suppose $(q,q_0)$ satisfies Hypothesis \ref{H3}. Then there is a number
$\tau$ such that $|\psi(z,0)|\geq|z|^g/3$ for all $z$ on the
semicircles given by $|z|=(2n\pi+\tau)/(2R)$ and $\Im(z)\leq 0$ and
sufficiently large integers $n$.
\end{lemma}

\begin{proof}
We only have to prove that the hypothesis of Lemma \ref{L5} is
satisfied. To this end define
$$\gamma(t)=K(t,0)\sum_{\ell=0}^g r_\ell(\xi(t)) z^{j-g}$$
and consider the integral
$$\int_0^{2R} \gamma(t) \exp(izt) dt.$$
According to Lemmas \ref{L5.2} and \ref{L5.3} this expression may be
integrated by parts $n+1$ times. Thus we obtain
\begin{align*}
z^{-g}\varphi(z,0)&=\sum_{j=0}^{n+1} (-1)^j \gamma^{(j)}(t)
 \frac{\e^{izt}}{(iz)^{j+1}}\Bigr|_0^{2R}
 +(-1)^n \int_0^{2R} \gamma^{(n+2)}(t)\frac{\e^{izt}}{(iz)^{n+2}}dt\\
 &=\sum_{j=0}^{n+1} \frac{\gamma^{(j)}(0)}{(-iz)^{j+1}}
 -\frac{\e^{2izR}}{(-iz)^{n+2}} \left(\gamma^{(n+1)}(2R)
 - \int_0^{2R} \gamma^{(n+2)}(t)\e^{iz(t-2R)}dt\right)
 \end{align*}
since $\gamma(2R)=...=\gamma^{(n)}(2R)=0$. However,
$$\gamma^{(n+1)}(2R)=-\frac1{2^{n+2}}(q-q_0)^{(n)}(R)
 \sum_{\ell=0}^g r_\ell(\xi(2R))z^{j-g}\neq0$$
for all sufficiently large $z$.

The Riemann-Lebesgue lemma gives that $\int_0^{2R}\gamma^{(n+2)}(t)
\e^{iz(t-2R)} dt$ tends to zero as $\Re(z)$ tends to infinity when
$\Im(z)$ is fixed. A closer look at its proof reveals that this is in
fact true uniformly in $\Im(z)$ as long as $\Im(z)$ is bounded above.
Hence there is a positive $X$ such that
$$\left|\int_0^{2R} \gamma^{(n+2)}(t)\e^{iz(t-2R)}dt\right|
 \leq \frac{|\gamma^{(n+1)}(2R)|}{12}$$
as long as $\Im(z)\leq 0$ and $|\Re(z)|\geq X$. It is also obvious that
$$\left|\sum_{j=0}^{n+1} \frac{\gamma^{(j)}(0)}{(-iz)^{j+1}}\right| \leq
 \frac16$$
if $|z|$ is sufficiently large.
\end{proof}

\subsection{The zeros of the Jost function determine the potential}
Suppose that $(q,q_0)$ satisfies Hypotheses \ref{H3} and let $\psi_0$
and $\psi$ be as before, in particular,
\begin{align}
\psi(z,x)&=\psi_0(z,x)+\int_0^{2R}K(t,x)\psi_0(z,t)dt,\label{170303.1}\\
\psi'(z,x)&=\psi_0'(z,x)+\int_0^{2R} K^{(0,1)}(t,x)\psi_0(z,t)dt.
\label{170303.2}
\end{align}
and
$$W(z)=\psi_0(z,x)\psi_0'(-z,x)-\psi_0'(z,x)\psi_0(-z,x).$$

Now define
$$s(z,t,x)=\frac1{W(z)}
 \left(\psi_0(-z,x)\psi_0(z,t)-\psi_0(z,x)\psi_0(-z,t)\right)$$
and $F(z,t,x)=-s(z,t,x)(q-q_0)(t)$. Then $\psi(z,\cdot)$ satisfies the
integral equation $\psi(z,x)=\psi_0(z,x)+\int_x^R F(z,t,x)\psi(z,t)dt$.
Therefore
$$F'(z,t,x)=M_0(z)F(z,t,x)-\frac{\psi_0(z,t)}{\psi_0(z,x)}(q-q_0)(t)$$
and
\begin{equation} \label{2606.1}
M(z)=M_0(z)-\frac1{\psi_0(z,0)\psi(z,0)} \int_0^R
 \psi_0(z,t)\psi(z,t)(q-q_0)(t)dt.
\end{equation}

\begin{thm} \label{example}
Suppose that $(q,q_0)$ satisfies Hypotheses \ref{H3} and that $q_0\in
\mathcal C_W$ for some odd polynomial $W$. Then also $q\in \mathcal
C_W$.

In particular, if none of the zeros of the Jost function
$\psi(\cdot,0)$ coincides with any of the zeros of $W$ then the zeros
of $\psi(\cdot,0)$ and their multiplicities determine $q$ uniquely.
\end{thm}

\begin{proof}
Theorem \ref{t3} proves the existence of a function $\psi$ satisfying
conditions (1) of Definition \ref{dcw}. Conditions (2) and (3) are
satisfied because they hold for $\psi_0(z,\cdot)$. Condition (4)
follows from equations \eqref{170303.1} and \eqref{170303.2}. In
particular, $\psi(\cdot,x)$ has growth order one since this is true for
$\psi_0(\cdot,x)$ and since $K(\cdot,x)$ is compactly supported.
Similarly, $\psi'(\cdot,x)$ has growth order one. As $z$ tends to
infinity along the imaginary axis $\psi_0(z,0)$ approaches $z^g$ while,
by Lemma \ref{L3a}, $\varphi(z,0)$ tends to zero. This implies
Condition (5).

We will now check condition (6) of the definition of the class
$\mathcal C_W$. Suppose first that $\Im(z)\geq 0$. For sufficiently
large $z$ we obtain from Lemma \ref{L3a} that
$$\frac12 |z|^g \leq |\psi(z,t)|, |\psi_0(z,t)| \leq \frac32|z|^g.$$
This and equation \eqref{2606.1} gives
$$|M(z)-M_0(z)| \leq 9 \|q-q_0\|_1.$$

To estimate $M(z)$ for $z$ in the lower halfplane note that the
Wronskian of $\psi(z,\cdot)$ and $\psi(-z,\cdot)$ satisfies
$$W(\psi(z,\cdot),\psi(-z,\cdot))=W(z).$$
Hence
$$M(z)=M(-z)-\frac{W(z)}{\psi(z,0)\psi(-z,0)}.$$
If $z$ is on the semicircles described in Lemma \ref{L7} then,
according to that Lemma, $|\psi(z,0)|\geq |z|^g/3$. This and the fact
that $|W(z)|\leq C |z|^{2g+1}$ imply that
$$|M(z)-M_0(-z)|
 \leq |M(-z)-M_0(-z)|+\frac{|W(z)|}{|\psi(z,0)\psi(-z,0)|}
 \leq 9 \|q-q_0\|_1+6 C |z|.$$

The last statement of the theorem is now simply an application of
Theorem~\ref{main}.
\end{proof}

\def\cprime{$'$} \def\cprime{$'$} \def\cprime{$'$}


\end{document}